\documentclass[a4paper,11pt]{article}
\usepackage{amsmath,amssymb,amsthm}
\usepackage[sort,authoryear]{natbib}

\usepackage{graphicx}

\newtheorem{proposition}{Proposition}
\newtheorem{theorem}{Theorem}

\newtheorem{lemma}{Lemma}

\theoremstyle{definition}
\newtheorem{example}{Example}

\numberwithin{equation}{section}

\begin{document}

\date{}
\title{On the asymptotic behavior of the solutions to the replicator equation}

\author{Georgy P Karev$^{1,}$\footnote{Corresponding author: tel.: +1 (301) 451-6722; e-mail: karev@ncbi.nlm.nih.gov}\,\,, Artem S Novozhilov$^{1,2}$\footnote{e-mail: anovozhilov@gmail.com}\,, Faina S Berezovskaya$^{3}$\footnote{e-mail: fberezovskaya@howard.edu}\\
\textit{\normalsize $^{1}$National Institutes of Health, 8600
Rockville Pike}\\ \textit{Bethesda MD 20894, USA}\\
\textit{\normalsize $^{2}$Moscow State University of Railway Engineering,}\\ \textit{Obraztsova 9, Moscow 127994, Russia}\\
\textit{\normalsize $^{3}$Howard University, 6-th Str. Washington DC 20059, USA}}

\maketitle

\begin{abstract} Selection systems and the corresponding replicator equations model the evolution of replicators with a high level of abstraction. In this paper we apply novel methods of analysis of selection systems to the replicator equations. To be suitable for the suggested algorithm the interaction matrix of the replicator equation should be transformed; in particular the standard singular value decomposition allows us to rewrite the replicator equation in a convenient form. The original $n$-dimensional problem is reduced to the analysis of asymptotic behavior of the solutions to the so-called escort system, which in some important cases can be of significantly smaller dimension than the original system. The Newton diagram methods are applied to study the asymptotic behavior of the solutions to the escort system, when interaction matrix has rank 1 or 2. A general replicator equation with the interaction matrix of rank 1 is fully analyzed; the conditions are provided when the asymptotic state is a polymorphic equilibrium. As an example of the system with the interaction matrix of rank 2 we consider the problem from [Adams, M.R. and Sornborger, A.T., J Math Biol, 54:357-384, 2007], for which we show, for arbitrary dimension of the system and under some suitable conditions, that generically one globally stable equilibrium exits on the 1-skeleton of the simplex.
\paragraph{\small Keywords:} replicator equation, selection system, singular value decomposition, Newton diagram, power asymptotes
\paragraph{\small AMS (MOS) subject classification:} 92B05, 92D15, 34C05, 34D05
\end{abstract}
\section{Introduction. Selection systems and replicator equations}\label{Sec1}
The evolution of replicators, which are the basic entities of the theory of natural selection, can be described with a high level of abstraction by means of the so-called selection systems or corresponding replicator equations, which, in their most generic form, can be written as follows.

Let us suppose that every individual of a population is characterized by its own value of parameter $\omega$, and $\omega$ takes its values from a measurable space $(\Omega,P)$  where $\Omega$ is the set of admissible parameter values and $P$ is a given measure. The parameter $\omega$ specifies an individual's invariant property and in most applications takes values in a finite set or in a domain of $n$-dimensional Euclidian space. Let us denote $l(t,\omega)$  the density of individuals with a given parameter value $\omega$  with respect to the measure $P$; the distribution of this parameter can be continuous or discrete, depending on the nature of the problem we seek to describe with our mathematical model.

An abstract selection system (or, in the author's terms, a system with inheritance) was studied in \cite{gorban2008a} (see also references to earlier work therein) where a general selection theorem was proven. Roughly speaking, one of the statements of the theorem is that an infinite-dimensional abstract system with inheritance ``tends'' in the course of time to a finite dimensional system (see \cite{gorban2008a} for the exact statements). This result justifies the special attention to selection systems with a discrete distribution of the parameter. The simplest example of a discrete distribution of parameter $\omega$ is given by interpretation of $\omega$ as merely an index of interacting subpopulations; in this case $l(t,\omega)$ is naturally to interpret as the size of the $\omega$-th subpopulation. To emphasize the discrete nature of the distribution of $\omega$ in some problems, we will use it as an index: $l_{\omega}(t)$ (or, more traditionally, $l_i(t)$, replacing $\omega$ with the index $i$).

If we denote the \textit{per capita} growth rate of $\omega$-th subpopulations as $F(t,\omega)$ and assume the overlapping generations and smoothness of $l(t,\omega)$ in $t$ for each fixed $\omega$, we obtain the abstract selection system (e.g., \cite{gorban2008a})
\begin{equation}\label{S1:Eq:1}
    \frac{\partial}{\partial t}\,l(t,\omega)=l(t,\omega)F(t,\omega),\quad l(t,\omega)\geq 0,
\end{equation}
where the initial condition $l(0,\omega)$ is given, and the growth rate, or fitness, $F(t,\omega)$ can depend, among other things, on the total population size $N(t)=\int_\Omega l(t,\omega)\,d\omega$ (where the integral is replaced with the sum if the distribution of $\omega$ is discrete). The exact form of $F(t,\omega)$ we will be working with is given below.

It is straightforward to infer that the frequencies of subpopulations, $$p(t,\omega)=\frac{l(t,\omega)}{N(t)}\,,$$ satisfy the replicator equation:
\begin{equation}\label{S1:Eq:2}
    \frac{\partial}{\partial t}\,p(t,\omega)=p(t,\omega)\left(F(t,\omega)-E_t[F]\right),
\end{equation}
where $E_t[F]=\int_\Omega F(t,\omega)p(t,\omega)\,d\omega$ denotes the mean fitness of the total population at the time $t$. The natural phase space of the replicator equation \eqref{S1:Eq:2} is given by $\{p(t,\omega)\colon p(t,\omega)\geq 0,\,\int_{\Omega}p(t,\omega)\,d\omega=1\}$, in the discrete case it is the simplex $S_{n}=\{p_i(t)\colon p_i(t)\geq 0,\,\sum_ip_i(t)=1\}$ (here and below we assume that generally there are $n$ interacting subpopulations, and the notation $\sum_i$ means $\sum_{i=1}^n$). We remark that the same replicator equation \eqref{S1:Eq:2} can be obtained for different selection systems \eqref{S1:Eq:1}, e.g., it is true if the growth rates in two selection systems \eqref{S1:Eq:1} differ by a function that depends only on the total population size; when passing in the opposite direction, from the replicator equation to the selection system we always choose the simplest one.

Naturally, equation \eqref{S1:Eq:2} should be supplemented with the equation
$$
\frac{d}{dt}N(t)=E_t[F]N(t),
$$
if the fitness $F(t,\omega)$ depends on $N(t)$.

The replicator equation \eqref{S1:Eq:2} comprises well-established biomathematical models in quite distinct evolutionary contexts, see \cite{hofbauer2003egd,hofbauer1998ega,schuster1983rd}. We survey these models briefly.

One of the first replicator equations was used, at least implicitly, by Ronald Fisher, John Haldane, and Sewall Wright to study the evolution of multiallelic one-locus gene frequencies under the force of natural selection in a sexually reproducing diploid population (for more information see \cite{hofbauer1998ega,hofbauer2003egd}). If a gene locus with $n$ alleles is considered, the frequency of the $i$-th allele is denoted as $p_i$, and the  Hardy--Weinberg equilibrium is assumed, then, in the usual way, the replicator equation is obtained:
\begin{equation}\label{S1:Eq:3}
\frac{d}{dt}p_i(t)=p_i(t)\left(\sum\nolimits_j m_{ij}p_j(t)-\sum\nolimits_{j,\,k}m_{jk}p_j(t)p_k(t)\right),\quad i=1,\ldots,n,
\end{equation}
where $m_{ij}$ is the Malthusian fitness of genotype with alleles $i$ and $j$. Given the assumptions, we have that in this case $m_{ij}=m_{ji}$, the fitness matrix $\textbf{M}=\{m_{ij}\}$ is symmetric.
In equation \eqref{S1:Eq:3} the intrinsic growth rate of the $i$-th allele depends linear on the frequencies of other alleles, in our notations,
$$
F_\omega(t)=\sum\nolimits_{j}m_{\omega j}p_j(t),\quad \omega=1,\ldots,n.
$$
If the fitnesses $m_{ij}$ are constant, then it is known that \eqref{S1:Eq:3} is a gradient system (\cite{svirezhev:fme,shahshahani1979nmf,hofbauer1998ega}), and the mean fitness of the population is monotonically increasing.

Sometimes it is natural to assume that the corresponding growth rates depend also on the population size and therefore include the density-dependent effects in the model, e.g., similar to the well-known Verhulst--Pearl logistic equation (e.g., \cite{ginzburg1977eas,charlesworth1971sdr}). One particular model with explicit birth and death terms, considered in~\cite{desharnais1983nsa}, has the form
\begin{equation}\label{S1:Eq:4}
m_{ij}=b_{ij}f(N)-d_{ij}g(N),
\end{equation}
where $b_{ij}$ and $d_{ij}$ are the per capita density-independent rates of recruitment and mortality, respectively, associated with the genotype $\{i,\,j\}$. Therefore to describe the evolution of the allele frequencies it is necessary to consider the following problem
\begin{equation}\label{S1:Eq:5}
\begin{split}
\frac{d}{dt}p_i&=p_i\left((b_i-E_t[\textbf{b}])f(N)+(E_t[\textbf{d}]-d_i)g(N)\right),\quad i=1,\ldots,n,\\
\frac{d}{dt}N&=N\left(E_t[\textbf{b}]f(N)-E_t[\textbf{d}]g(N)\right),
\end{split}
\end{equation}
where $b_i=\sum\nolimits_jb_{ij}p_j,\,d_i=\sum\nolimits_jd_{ij}p_j,\,E_t[\textbf{b}]=\sum\nolimits_ib_{i}p_i,\,E_t[\textbf{d}]=\sum\nolimits_id_{i}p_i$, and the explicit dependence on $t$ for $p_i$ and $N$ was suppressed for simplicity. System \eqref{S1:Eq:5} can be replaced with the following selection system
\begin{equation*}
\begin{split}
\frac{d}{dt}l_i&=l_i\left(b_if(N)-d_ig(N)\right),\quad i=1,\ldots,n,\\
N(t)&=\sum\nolimits_i l_i(t).
\end{split}
\end{equation*}

Another example of the replicator equation is given by the equation for the frequencies of the pure strategies in the population within the framework of the evolutionary game theory (see \cite{taylor1978ess,hofbauer1998ega,hofbauer2003egd}). If it is assumed that $a_{ij}$ denotes the payoff of the player with pure strategy $i$ against the player with pure strategy $j$, then the dynamics of the frequencies of the players in the population are given by
\begin{equation}\label{S1:Eq:6}
\frac{d}{dt}p_i(t)=p_i(t)\left(\sum\nolimits_j a_{ij}p_j(t)-\sum\nolimits_{j,\,k}a_{jk}p_j(t)p_k(t)\right),\quad i=1,\ldots,n,
\end{equation}
where, contrast to \eqref{S1:Eq:3}, the matrix $\textbf{A}$ is an arbitrary real $n\times n$ matrix. In the case of a continuum of pure strategies an analogue to \eqref{S1:Eq:2} is obtained (\cite{bomze1990dae,hofbauer2003egd}). System \eqref{S1:Eq:3} is a particular case of \eqref{S1:Eq:6} with a symmetric $\textbf{A}$, such matrices describe partnership games. The selection system, corresponding to \eqref{S1:Eq:6},  has the form
$$
\frac{d}{dt}l_i(t)=l_i(t)\left(\sum\nolimits_j a_{ij}\frac{l_j(t)}{\sum\nolimits_m l_m(t)}\right).
$$

Another well studied particular case of \eqref{S1:Eq:6} is the so-called hypercycle equation, which is given by setting $a_{ij}=k_i$ if $j=i-1$ and $a_{ij}=0$ otherwise. The hypercycle equation describes a catalytic loop of self-replicating macromolecules, each promoting replication of another type (\cite{eigenshuster}). Completing the list of application of the replicator equation \eqref{S1:Eq:2} we note that the classical equations of Volterra describing dynamics of interacting populations can be transformed into the form \eqref{S1:Eq:2} by means of an invertible change of variables (see \cite{hofbauer1998ega}). However, we remark that the methods, described in the present text, are better to apply directly to the Volterra systems.

Effective methods of analysis of selection systems \eqref{S1:Eq:1} were developed recently (see \cite{karev2009} and references therein) for particular form of the fitness function $F(t,\omega)$. Here we present the applications of these methods to the selection system \eqref{S1:Eq:1} and, consequently, to the replicator equation \eqref{S1:Eq:2}.  It turns out that some of the systems obeying the replicator equations can be effectively analyzed and solved even for large~$n$.

Our paper organized as follows. In the next section we present an algorithm, which allows us to replace a given selection system with an equivalent problem. This equivalent problem in some particular cases can be of significantly lower dimension than the original system, and this is the case when the suggested methods should be taken advantage of. Section 3 is devoted to the methods how to transform a given replicator equation so that the methods of Section 2 can be directly applicable. In Section 4 we present a non-trivial application of the proposed technique to the replicator equation, which is supposed to model the evolution of sensory systems, and obtain a general proof, under some suitable conditions, of a conjecture, which was proved only for particular cases in the original study \cite{adams2007acc}. The last section is devoted to conclusions, and Appendix contains some auxiliary facts.

\section{How to solve selection systems}\label{Sec2}
Here we present an algorithm that allows us to formally write down an explicit solution to the selection system \eqref{S1:Eq:1}, which we rewrite here for convenience,
\begin{equation*}
    \frac{\partial}{\partial t}\,l(t,\omega)=l(t,\omega)F(t,\omega),\tag{\ref{S1:Eq:1}}\quad l(t,\omega)\geq 0,
\end{equation*}
when the fitness $F(t,\omega)$ has the following special form:
\begin{equation}\label{S2:Eq:1}
F(t,\omega)=\sum\nolimits_{i=1}^{m_1}u_i(t,G_i)\varphi_i(\omega)+\sum\nolimits_{j=1}^{m_2}v_j(t,H_j)\psi_j(\omega),
\end{equation}
where $G_i,\,H_j$ are the so-called regulators
\begin{equation}\label{S2:Eq:2}
\begin{split}
G_i(t)&=\int_{\Omega}g_i(\omega)l(t,\omega)\,d\omega=N(t)E_t[g_i],\quad i=1,\ldots,m_1,\\
H_j(t)&=\int_{\Omega}h_j(\omega)p(t,\omega)\,d\omega=E_t[h_i],\quad j=1,\ldots,m_2,
\end{split}
\end{equation}
$u_i,\,v_j,\,g_i,\,h_j,\,\varphi_i$ and $\psi_j$ are given functions, $m_1,\,m_2\geq 0$ are constants, and $p(t,\omega)=l(t,\omega)/N(t)$. The probability density function $p(t,\omega)$ solves the replicator equation \eqref{S1:Eq:2}.

In applications functions $\varphi_i(\omega),\,\psi_j(\omega)$ can be interpreted, for instance, as particular phenotype traits that characterize an individual with the parameter value $\omega$; $u_i(t,G_i)$ and $v_j(t,H_j)$ then describe the contribution of the corresponding phenotype traits to the fitness (mean number of descendants per individual) at the time moment $t$ provided the values of regulators $G_i$ and $H_j$. Note that the traits $\{\varphi_i\}$ correspond to the density-dependent regulators $G_i$, whereas the traits $\{\psi_j\}$ correspond to the frequency-dependent regulators $H_j$. We divide the regulators into these two group for convenience, although it should be clear that the theory could be written only for $G_i$.

This particular form of the selection system \eqref{S1:Eq:1},\eqref{S2:Eq:1} comprises many meaningful mathematical models (see \cite{karev2009} for the general theory and, e.g., \cite{karev2003imt,karev2005dip,novozhilov2004agp,Novozhilov2008,karev2006nov} for various applications).  Model \eqref{S1:Eq:1},\eqref{S2:Eq:1},\eqref{S2:Eq:2} defines, in general, a complex transformation of the initial distribution $l(0,\omega)$. The remarkable fact is that model \eqref{S1:Eq:1},\eqref{S2:Eq:1} can be reduced to an equivalent system of ordinary differential equations (ODEs). Here we present only the algorithm; the proofs can be found in~\cite{karev2009}.

We introduce the functional on the space of measurable functions of the parameter $\omega$:
\begin{equation}\label{S2:Eq:3}
\begin{split}
M(z;\boldsymbol\lambda,\,\boldsymbol\delta)&=\int_{\Omega}z(\omega)\exp\left\{\sum\nolimits_{i=1}^{m_1}\lambda_i\varphi_i(\omega)+\sum\nolimits_{j=1}^{m_2}\delta_j\psi_j(\omega)\right\}p(0,\omega)\,d\omega=\\
&=E_0\left[z\exp\left\{\sum\nolimits_{i=1}^{m_1}\lambda_i\varphi_i+\sum\nolimits_{j=1}^{m_2}\delta_j\psi_j\right\}\right],
\end{split}
\end{equation}
where $p(0,\omega)=l(0,\omega)/N(0)$, and $\boldsymbol\lambda=(\lambda_1,\ldots,\lambda_{m_1})$, $\boldsymbol\delta=(\delta_1,\ldots,\delta_{m_2})$.

Next we write down \textit{the {escort} system} of ordinary differential equations:
\begin{equation}\label{S2:Eq:4}
\begin{split}
\frac{d}{dt}q_i(t)&=u_i(t,N(0)M(g_i;\textbf{q}(t),\,\textbf{s}(t))),\quad q_i(0)=0,\quad i=1,\ldots,m_1,\\
\frac{d}{dt}s_j(t)&=v_j(t,M(h_j;q(t),\,s(t))/M(1;\,\textbf{q}(t),\,\textbf{s}(t))),\quad s_j(0)=0,\quad j=1,\ldots,m_2,
\end{split}
\end{equation}
where $\textbf{q}(t)=(q_1(t),\ldots,q_{m_1}(t))$ and $\textbf{s}(t)=(s_1(t),\ldots,s_{m_2}(t))$.

Using the functional \eqref{S2:Eq:3} and the solutions to \eqref{S2:Eq:4} we can write the solution to the selection system \eqref{S1:Eq:1},\eqref{S2:Eq:1} as
\begin{equation}\label{S2:Eq:5}
l(t,\omega)=l(0,\omega)K(t,\omega),
\end{equation}
where
\begin{equation}\label{S2:Eq:6}
K(t,\omega)=\exp\left\{\sum\nolimits_{i=1}^{m_1}q_i(t)\varphi_i(\omega)+\sum\nolimits_{j=1}^{m_2}s_j(t)\psi_j(\omega)\right\}.
\end{equation}
We also have that the total population size is equal to
\begin{equation}\label{S2:Eq:7}
N(t)=N(0)M(1;\,\textbf{q}(t),\,\textbf{s}(t));
\end{equation}
the values of regulators are given by
\begin{equation}\label{S2:Eq:8}
\begin{split}
G_i(t)&=N(0)M(g_i;\,\textbf{q}(t),\,\textbf{s}(t)),\quad i=1,\ldots,m_1,\\
H_j(t)&=\frac{M(h_j;\,\textbf{q}(t),\,\textbf{s}(t))}{M(1;\,\textbf{q}(t),\,\textbf{s}(t))}\,,\quad j=1,\ldots,m_2;
\end{split}
\end{equation}
and the current probability density function of the parameter distribution can be presented in the explicit form as
\begin{equation}\label{S2:Eq:9}
p(t,\omega)=p(0,\omega)\,\frac{K(t,\omega)}{M(1;\,\textbf{q}(t),\,\textbf{s}(t))}=p(0,\omega)\,\frac{K(t, \omega)}{E_t\left[K(t,\cdot)\right]}\,.
\end{equation}

Formula \eqref{S2:Eq:9} is the central result of the theory; it gives the solution to the replicator equation~\eqref{S1:Eq:2} and allows us to compute all the statistical characteristics of the underlying parameter distribution in the self-regulated selection systems of the form \eqref{S1:Eq:1},\eqref{S2:Eq:1}.

The major technical tool in the considered approach is the functional $M(z;\,\boldsymbol\lambda,\,\boldsymbol\delta)$, which is formally well-defined for any given initial distribution $p(0,\omega)$; in practice, however, it might be difficult to evaluate $M(z;\,\boldsymbol\lambda,\,\boldsymbol\delta)$ for particular functions $z(\omega)$. A significant simplification is achieved if fitness $F(t,\omega)$ depends only on the regulators of the following form: $N(t),\,E_t[\varphi_i],$ or $N(t)E_t[\varphi_i]$. In this case it is straightforward to see that instead of the general functional $M(z;\,\boldsymbol\lambda,\,\boldsymbol\delta)$ we can use the moment generating function (mgf) of the initial distribution $p(0,\omega)$, which is defined as
$$
M_0(\boldsymbol\delta)=E_0\left[\exp\left\{\sum\nolimits_{i=1}^{m_1}\delta_i\varphi_i\right\}\right].
$$
Indeed,
\begin{equation}\label{S2:Eq:10}
\begin{split}
M(1;\,\textbf{q}(t))&=E_0\left[\exp\left\{\sum\nolimits_{i=1}^{m_1}q_i(t)\varphi_i\right\}\right]=M_0(\textbf{q}(t)),\\
M(\varphi_k;\textbf{q}(t))&=E_0\left[\varphi_k\exp\left\{\sum\nolimits_{i=1}^{m_1}q_i(t)\varphi_i\right\}\right]=\frac{\partial}{\partial \varphi_k}\,M_0(\textbf{q}(t)).
\end{split}
\end{equation}
The same holds for the frequency dependent regulators.

Using \eqref{S2:Eq:10} the right hand side of \eqref{S2:Eq:4} can be rewritten in terms of the moment generating function of the initial distribution, which is supposed to be given. Remark that the moment generating functions are known for many important probability density functions.

As a simple example we consider the following
\begin{example}\label{Examp1}
Let the Malthusian fitnesses be multiplicative, i.e., $m_{ij}=m_im_j$ for the given vector $\textbf{m}=(m_1,\ldots,m_n),$ where $m_i\neq m_j$ for any $i\neq j$. The asymptotic outcome of the dynamics is well known and trivial: all alleles but the one with the highest fitness $m_{max}$ are lost; here, additionally to this known result,  we obtain also a simple expression that can be used to compute time-dependent behavior. System \eqref{S1:Eq:3} for the multiplicative fitness can be rewritten as
\begin{equation}\label{S3:Eq:1}
\frac{d}{dt}p_i(t)=p_i\left(m_iE_t[\textbf{m}]-(E_t[\textbf{m}])^2\right),\quad i=1,\ldots,n,
\end{equation}
where $E_t[\textbf{m}]=\sum\nolimits_{j=1}^nm_jp_j(t)$. The following selection system corresponds to replicator equation \eqref{S3:Eq:1}:
\begin{equation}\label{S3:Eq:2}
\frac{d}{dt}l_i(t)=l_i(t)m_iE_t[\textbf{m}],\quad i=1,\ldots,n,
\end{equation}
and belongs to the class \eqref{S1:Eq:1}, \eqref{S2:Eq:1}.

As before, denote $M_0(\lambda)$ the moment generating function of the initial distribution $p_i(0)$, $M_0(\lambda)=\sum\nolimits_{i}\exp\{\lambda m_i\}p_i(0)=E_0[\exp\{\lambda \textbf{m}\}]$. The escort system \eqref{S2:Eq:4} consists only of one equation:
\begin{equation}\label{S3:Eq:3}
\frac{d}{dt}s(t)=\frac{1}{M_0(s(t))}\,\frac{d}{ds}M_0(s(t))=\frac{d}{ds}\ln M_0(s(t)),\quad s(0)=0.
\end{equation}
The solution for the frequencies is given, according to \eqref{S2:Eq:9}, by
\begin{equation}\label{S3:Eq:4}
p_i(t)=p_i(0)\,\frac{\exp\{m_is(t)\}}{E_0[K(t,\cdot)]}\,,\quad E_0[K(t,\cdot)]=\sum\nolimits_{j}\exp\{s(t)m_j\}p_i^0\,,
\end{equation}
where $p_i^0$ denote the initial conditions, $p_i^0=p_i(0)>0$ for any $i$.

From \eqref{S3:Eq:3}, and using the change of the variable $s(t)=-\ln u(t)$, we obtain the equation for the new variable $u$
$$
\dot{u}=-u\frac{\sum\nolimits_im_ip_i^0u^{d_i}}{\sum\nolimits_ip_i^0u^{d_i}},
$$
where $d_i=m_{max}-m_i$, which implies that one of $d_i=0$. From the last equation it follows that $u(t)\to0$ as $t\to\infty$, which yields that $s(t)\to\infty$. Using the last fact and the solution to \eqref{S3:Eq:3} we have
$$
\frac{p_i(t)}{p_j(t)}=\frac{p_i^0}{p_j^0}\exp\{s(t)(m_i-m_j)\}\to\infty
$$
if $m_i>m_j$, which is possible only if $p_j(t)\to 0$ due to the constraint $\sum\nolimits_i p_i(t)=1$.

The major advantage of the considered approach is that if one needs the time-dependent behavior of system \eqref{S3:Eq:1} then, instead of solving $n$ differential equation it is suffice to solve only one differential equation \eqref{S3:Eq:3} for the auxiliary variable $s(t)$. In a similar vein the case of additive fitness $m_{ij}=m_i+m_j$ can be analyzed.
\end{example}

\section{Reduction of a general replicator equation by means of matrix decompositions}\label{Sec3}
It is obvious that only in exceptional cases, as for the system \eqref{S3:Eq:1}, the theory of Section \ref{Sec2} can be applied to the replicator equation \eqref{S1:Eq:6} directly. In the general case we need to rewrite the interaction matrix $\textbf{A}$ from the equation \eqref{S1:Eq:6} in the form, suitable for the described technique. In this section we propose a method to apply general technique of the analysis of the selection system \eqref{S1:Eq:1} to the replicator equation.

We start with a symmetric matrix $\textbf{A}$.

\subsection{Spectral decomposition}\label{Examp3} Let us consider again the equation for the allele frequencies in diploid population \eqref{S1:Eq:3} with a constant matrix $\textbf{M}$. According to the interpretation of this equation, matrix $\textbf{M}$ is symmetric, $m_{ij}=m_{ji}$. Any symmetric real matrix $\textbf{M}$ can be presented in the form (e.g., \cite{ortega1987mts})
\begin{equation}\label{S3:Eq:8}
    \textbf{M}=\lambda_1\textbf{h}_1\textbf{h}_1^\tau+\lambda_2\textbf{h}_2\textbf{h}_2^\tau+\ldots+\lambda_k\textbf{h}_k\textbf{h}_k^\tau,
\end{equation}
where $\lambda_i,\,i=1,\ldots,k$ are the real eigenvalues of $\textbf{M}$, $k$ is the rank of $\textbf{M}$, $\textbf{h}_i,\,i=1,\ldots,k$ are the corresponding right eigenvectors that satisfy $\textbf{h}_i^\tau\textbf{h}_i=1,\,\textbf{h}_i^\tau\textbf{h}_j=0,\,i\neq j$, and $\tau$ denotes transposition. The form \eqref{S3:Eq:8} is the spectral decomposition of $\textbf{M}$. If we denote the $j$-th element of the $i$-th eigenvector $\textbf{h}_i$ as $h_{ji}$ then each element of $\textbf{M}$ has the form
$$
m_{ij}=\lambda_1h_{i1}h_{j1}+\lambda_2h_{i2}h_{j2}+\ldots+\lambda_kh_{ik}h_{jk}.
$$
According to the last equality, system \eqref{S1:Eq:3} takes the form
\begin{equation}\label{S3:Eq:9}
    \frac{d}{dt}p_i(t)=p_i(t)\left(\sum\nolimits_{j=1}^k\lambda_jh_{ij}E_t[\textbf{h}_j]-\sum\nolimits_{j=1}^k\lambda_j(E_t[\textbf{h}_j])^2\right),\quad i=1,\ldots,n,
\end{equation}
and hence belongs to the class of selection systems \eqref{S1:Eq:1} with fitness \eqref{S2:Eq:1}.

Consider the mgf of the initial distribution of $\textbf{h}_i$:
$$
M_0(\boldsymbol\delta)=\sum\nolimits_{i=1}^np_i^0\exp\left\{\sum\nolimits_{m=1}^k\delta_mh_{im}\right\}.
$$
The last expression yields the following escort system:
\begin{equation}\label{S3:Eq:10}
\begin{split}
    \frac{d}{dt}s_j(t)&=\lambda_j\frac{E_0[\textbf{h}_j\exp\{\sum\nolimits_{m=1}^ks_m(t)\textbf{h}_m\}]}{E_0[\exp\{\sum\nolimits_{m=1}^ks_m(t)\textbf{h}_m\}]}\\
                      &= \lambda_j   \frac{\sum\nolimits_{i=1}^np_i^0h_{ij}\sum\nolimits_{m=1}^ks_m(t)h_{im}}{\sum\nolimits_{i=1}^np_i^0\sum\nolimits_{m=1}^ks_m(t)h_{im}}\,,\quad j=1,\ldots,k.
\end{split}
\end{equation}
System \eqref{S3:Eq:10} can be rewritten in the compact form
\begin{equation}\label{S3:Eq:11}
    \frac{d}{dt}s_j=\lambda_j\frac{\partial}{\partial s_j}\ln M_0(\textbf{s}),\quad j=1,\ldots,k.
\end{equation}
Hence the solution to system \eqref{S1:Eq:3} is
$$
p_i(t)=p_i(0)\frac{K(t,i)}{E_0[K(t,\cdot)]}\,,\quad K(t,i)=\exp\left\{\sum\nolimits_{j=1}^ks_j(t)h_{ij}\right\},\quad E_0[K(t,\cdot)]=M_0(\textbf{s}(t)).
$$

To sum it up, we replace $n$-dimensional problem \eqref{S1:Eq:3} with $k$-dimensional system \eqref{S3:Eq:10}. We point out that the latter system can be no easier to solve than the original one. The suggested approach is beneficial only when $k\ll n$ (here we only speak of finite dimensional systems \eqref{S1:Eq:3}).


The spectral decomposition approach as in the previous example allows us to reduce the original problem \eqref{S1:Eq:6} to the system in the form \eqref{S1:Eq:1}, \eqref{S2:Eq:1} only for symmetric $\textbf{A}$. In the general case with an arbitrary real matrix $\textbf{A}$ we can apply the singular value decomposition (e.g., \cite{jolliffe2002pca}).

\subsection{Singular value decomposition}
Well known that, given an arbitrary matrix $\textbf{A}$ of dimension $n\times n$, $\textbf{A}$ can be written as
\begin{equation}\label{S4:Eq:1}
\textbf{A}=\textbf{U}\boldsymbol\Sigma \textbf{X}^\tau,
\end{equation}
where $\textbf{U},\,\textbf{X}$ are $n\times k$ matrices, each of which has orthonormal columns so that $\textbf{U}^\tau \textbf{U}=\textbf{X}^\tau \textbf{X}=\textbf{I}_k$, where $\textbf{I}_k$ is the identity $k\times k$ matrix; $\boldsymbol\Sigma$ is a $k\times k$ diagonal matrix with non-negative elements $\sigma_1\geq\sigma_2\geq\ldots\geq\sigma_k>0$ on the main diagonal; and $k$ is the rank of $\textbf{A}$ (see, e.g., \cite{jolliffe2002pca}). The representation \eqref{S4:Eq:1} is called singular value decomposition (SVD). Singular values $\sigma_j,\,i=1,\ldots,k$ are the square roots of the eigenvalues of $\textbf{AA}^\tau$ (or $\textbf{A}^\tau \textbf{A}$), which are given by $\sigma_1^2\geq\sigma_2^2\geq\ldots\geq\sigma_k^2>0$; columns of $\textbf{U}$ and $\textbf{X}$ are the right eigenvectors of $\textbf{AA}^\tau$ and $\textbf{A}^\tau \textbf{A}$ respectively; each column corresponds to its own $\sigma_j^2$.

From \eqref{S4:Eq:1} it follows that
$$
a_{ij}=\sum\nolimits_{m=1}^ku_{im}\sigma_m x_{jm},
$$
where $u_{im}$, $x_{jm}$ are the elements of $\textbf{U}$ and $\textbf{X}^\tau$ respectively. Using this representation, any matrix $\textbf{A}$ in \eqref{S1:Eq:6} can be written such that the fitness $F(t,\omega)$ in \eqref{S1:Eq:2} is in particular form \eqref{S2:Eq:1}. That is, we obtain
\begin{equation}\label{S4:Eq:2}
    \frac{d}{dt}p_i(t)=p_i(t)\left(\sum\nolimits_{j=1}^ku_{ij}\sigma_jE_t[\textbf{x}_j]-\sum\nolimits_{j=1}^k\sigma_j E_t[\textbf{u}_j]E_t[\textbf{x}_j]\right),
\end{equation}
where $\textbf{u}_j$, $\textbf{x}_j$ denote the $j$-th columns of matrices $\textbf{U}$ and $\textbf{X}^\tau$ respectively.

Singular value decomposition enables us to split elements $a_{ij}$ into parts
\begin{equation}\label{S4:Eq:3}
u_{im}\sigma_m x_{jm},\quad m=1,\ldots,k.
\end{equation}
If only $q<k$ such parts are retained, then the expression
$$
{}_q\tilde{a}_{ij}=\sum\nolimits_{m=1}^qu_{im}\sigma_m x_{jm}
$$
provides an approximation to $a_{ij}$ in a sense that ${}_q\tilde{a}_{ij}$ gives the best possible rank $q$ approximation to $a_{ij}$ (the proof can be found in \cite{gabriel1978lsa}) when
$$
\sum\nolimits_i\sum\nolimits_j({}_qa_{ij}-a_{ij})^2
$$
is minimized with respect to any matrix $({}_qa_{ij})$ of rank $k$.

The last point bears a close relationship with the principal component analysis (\cite{jolliffe2002pca}). Let us assume that matrix $\textbf{A}$ represents a table of $n$ observations of $m$ random variables. The usual way to reduce the dimensionality of this data-set, while retaining as much as possible variation presented in it, is to apply the principal component analysis. It can be shown that parts \eqref{S4:Eq:3} give the contribution of the corresponding principal components to the data-set; retaining only $q$ parts corresponds to retaining $q$ first principal components.

We remark that the singular value decomposition essentially depends on the choice of the scalar product (see, e.g., Ch.~5 in \cite{gorban2005invariant}). In the case of high- or infinite- dimension models we can, taking a finite number of components, reduce the original model to a finite dimensional selection system with the fitness in the particular form \eqref{S2:Eq:1}. In what follows we study only finite dimensional models of the form (6) with the standard scalar product. Note that if matrix $\textbf{A}$ has the rank $k$, then the escort system \eqref{S2:Eq:4} is $k$-dimensional.

It is reasonable to assume that matrix $\textbf{A}$ in applications is known only approximately. Retaining $q$ first principle components would correspond to the approximation of the matrix $\textbf{A}$ with the best possible matrix of rank $q$; from the standpoint of the theory presented in Section \ref{Sec2}, such approximation is useful because the dimension of the escort system is reduced to~$q$.

Let us rewrite system \eqref{S1:Eq:6} in the form
\begin{equation}\label{t1:1}
    \frac{d}{dt}p_i=p_i((\textbf{A}\textbf{p})_i-\textbf{p}^{\tau}\textbf{Ap}).
\end{equation}
The stationary points are found as the solutions of
$$
\textbf{Ap}=(\textbf{p}^{\tau}\textbf{Ap})\textbf{1}_n,
$$
where $\textbf{1}_n=(1,1,\ldots,1)^{\tau}$. Using \eqref{S4:Eq:1} the last system can be rewritten as
\begin{equation}\label{t1:2}
    \sigma_1E_t[\textbf{x}_1]\textbf{u}_1+\ldots+\sigma_kE_t[\textbf{x}_k]\textbf{u}_k-(\textbf{p}^{\tau}\textbf{Ap})\textbf{1}_n=0,
\end{equation}
which shows that in general an isolated polymorphic equilibrium can exists only when $k=n$, or when vectors $\textbf{u}_1,\,\ldots,\,\textbf{u}_k,\,\textbf{1}_n$ are linear dependent (cf. with the proof of similar conjecture in \cite{Sornborger2008}). On the other hand, when $k<n$ it is possible to have manifolds of non-isolated equilibria for which $E_t[\textbf{x}_1]=\ldots =E_t[\textbf{x}_k]=0$ (see Section \ref{sec31} for a particular example). Therefore, approximation of the original system with a matrix of smaller rank can yield either lost of information on isolated polymorphic equilibria, or appearance of manifolds of non-isolated equilibria.

On the other hand, it is reasonable to expect that in applications matrix $\textbf{A}$ can have exactly rank $k$, whereas its estimate $\tilde{\textbf{A}}$, which is known to the researcher, can have the rank $n$ due to, e.g., noise effects. In this case the reduction technique does not loose any information, and, additionally, discard the fictional information, which appears in the matrix $\textbf{A}$ thanks to the estimate errors. Again, using analogies with the principal component analysis, we can use known technique to infer the dimension (the rank of the matrix) that contains the principal information (for a review article how to determine the number of significant principal components see, e.g.,~\cite{cangelosi2007crp}).

It is interesting to note that for the case of a symmetric $\textbf{A}$ the escort system is a gradient system.
\begin{example}[Partnership games and gradients]
In Example \ref{Examp3} we already considered the replicator equation with symmetric matrix using the spectral decomposition. Here we apply SVD to such systems.

Let $\textbf{A}^\tau=\textbf{A}$, then \eqref{S4:Eq:1} becomes $\textbf{A}=\textbf{U}\boldsymbol\Sigma \textbf{U}^\tau$ since $\textbf{AA}^\tau=\textbf{A}^\tau \textbf{A}$, and each element of $\textbf{A}$ has the form
$$
a_{ij}=\sum\nolimits_{m=1}^ku_{im}\sigma_mu_{jm}.
$$
Denote $\tilde{\textbf{u}}_i=\sigma^{1/2}\textbf{u}_i$, where $\textbf{u}_i$ is the $i$-th column of $\textbf{U}$. Then we have that $\textbf{A}=\tilde{\textbf{U}}\tilde{\textbf{U}}^\tau$, and $a_{ij}=\sum\nolimits_{m=1}^k\tilde{u}_{im}\tilde{u}_{jm}$. Finally, the replicator equation takes the form
$$
\frac{d}{dt}p_i(t)=p_i(t)\left(\sum\nolimits_{j=1}^k\tilde{u}_{ij}E_t[\tilde{\textbf{u}}_j]-\sum\nolimits_{j=1}^k(E_t[\tilde{\textbf{u}}_j])^2\right),\quad i=1,\ldots,n,
$$
with the corresponding selection system
\begin{equation}\label{S4:Eq:4}
\frac{d}{dt}l_i(t)=l_i(t)\sum\nolimits_{j=1}^k\tilde{u}_{ij}E_t[\tilde{\textbf{u}}_j],\quad i=1,\ldots,n,
\end{equation}
which is as required by \eqref{S2:Eq:1}. As before denote $M_0(\boldsymbol\delta)$ the moment generating function of the initial distribution of the elements of $\tilde{\textbf{u}}_j$. The escort system now reads
\begin{equation}\label{S4:Eq:5}
\frac{d}{dt}s_j=\frac{\partial}{\partial s_j}\ln M_0(\textbf{s}),\quad s_j(0)=0,\quad j=1,\ldots,n,
\end{equation}
or simply
$$
\dot{\textbf{s}}=\nabla \ln M_0(\textbf{s}),\quad s(0)=0,
$$
which is a gradient system with the potential $-\ln M_0(s)$ in the usual space with the standard Euclidian metric.
\end{example}

\subsection{Analysis of the replicator equation having the matrix of rank 1}\label{sec31}

To conclude this section we consider the problem of finding all possible asymptotic states of the replicator equation with the interaction matrix that has rank 1 (note that Example \ref{Examp1} is a particular case of this problem).

According to SVD any matrix of rank 1 can be presented as $\textbf{A}=\textbf{ab}^{\tau}$, where $\textbf{a}$ and $\textbf{b}$ are vectors.

It is a simple matter to determine the asymptotic states in the replicator system with such matrix $\textbf{A}$ in the case $\textbf{b}\geq0$.

The escort system reads
\begin{equation}\label{eq1}
    \dot{s}=\frac{\sum\nolimits_ib_ip_i^0\exp\{a_is\}}{\sum\nolimits_ip_i^0\exp\{a_is\}}\,,
\end{equation}
where all $p_i^0>0$, and the initial condition $s(0)=0$.

Using the change of the variable $u=\exp\{-s\}$ we obtain that
\begin{equation}\label{eq2}
    \dot{u}=-u\frac{\sum\nolimits_ib_ip_i^0u^{d_i}}{\sum\nolimits_ip_i^0u^{d_i}}\,,
\end{equation}
where $d_i=\max_i\{a_i\}-a_i$. Note that at least one $d_i=0$, so that the previous equation can be rewritten as
\begin{equation}\label{eq3}
    \dot{u}=-u\frac{\sum\nolimits_{i\in I_k} b_ip_i^0+\sum\nolimits_{i \notin I_k}b_ip_i^0u^{d_i}}{\sum\nolimits_{i \in I_k}p_i^0+\sum\nolimits_{i \notin I_k}p_i^0u^{d_i}}\,,
\end{equation}
where $I_k=\{i\colon a_i=\max_j\{a_j\}\}$, and all $d_i>0$ if ${i \notin I_k}$. From \eqref{eq3} and if $\textbf{b}\geq 0$ it follows that the origin is a grobally asymptotically stable equilibrium and, when $t\to\infty$
$$
u(t)\to\exp\{-\lambda t\},\quad \lambda=\frac{\sum\nolimits_{i \in I_k}b_ip_i^0}{\sum\nolimits_{i \in I_k}p_i^0}\,,
$$
which means that $s(t)/t\to\lambda$ as $t\to\infty$.

We have that
$$
p_j(t)=\frac{p_j^0\exp\{a_js(t)\}}{\sum\nolimits_ip_i^0\exp\{a_is(t)\}}=\frac{p_j^0u^{d_j}}{\sum\nolimits_ip_i^0u^{d_i}}\,,
$$
which yields that
\begin{equation}\label{tod:1}
p_j(t)\to\frac{p_j^0}{\sum\nolimits_{i \in I_k}p_i^0}\quad \mbox{if } i\in I_k,
\end{equation}
and $p_j(t)\to0$ otherwise, which is generalization of the result in Example \ref{Examp1}. Note that if $\textbf{b}\leq 0$ a similar result is valid if we denote $I_k=\{i\colon a_i=\min_j \{a_j\}\}$.

To analyze the general case, when elements of $\textbf{b}$ can have arbitrary signs, we introduce two parameters: $\xi_1=\sum\nolimits_{i\in I_k}b_ip_i^0,\,\xi_2=\sum\nolimits_i b_ip_i^0=E_0[\textbf{b}]$.

If $\xi_1>0$ and $\xi_2>0$ then, due to continuity of the right hand side of \eqref{eq3}, equation \eqref{eq3} may possess zero or even number of equilibria $\{\hat{u}_k\}$. Using the fact that $\hat{u}=0$ is a stable equilibrium and recalling that the initial condition is $u(0)=1$, we obtain that $u(t)\to 0$ or $u(t)\to \hat{u}^*$ when $t\to\infty$, where $\hat{u}^*$ is the closest equilibrium of \eqref{eq2} to $u=1$ belonging to $(0,1)$.

In the case $\xi_1<0,\,\xi_2>0$ there is always equilibrium $\hat{u}^*\in(0,1)$, and $\hat{u}=0$ is unstable, which means that $u(t)\to\hat{u}^*$ when $t\to\infty$.

In the case $\xi_1>0,\,\xi_2<0$ it is possible that $u(t)\to\infty$ when $t\to\infty$ if there are no equilibria of \eqref{eq3} when $u\in(1,\infty)$, or $u(t)\to \hat{U}^*$, where $\hat{U}^*$ is the closest equilibrium of \eqref{eq3} belonging to $(1,\infty)$. The case $\xi_1<0,\,\xi_2<0$ is analogous to the previous one.

Therefore, we showed that $u(t)$ can tend to $0$, $\hat{u}$ or $\infty$ when $t\to\infty$. Using this fact, the explicit expression for frequencies, and auxiliary notation $J=\{j\colon a_j=\min_i\{a_i\}\}$, we obtain
\begin{proposition}\label{prop2}
Three types of asymptotic behavior of the solutions $p_i(t)$ to the replicator equation \eqref{S1:Eq:6} are possible, if the interaction matrix of the replicator equation has rank 1:

\emph{1)} If $u(t)\to 0$ as $t\to\infty$ then
\begin{equation*}
p_j(t)\to\frac{p_j^0}{\sum\nolimits_{i \in I_k}p_i^0}\quad \mbox{if } j\in I_k,
\end{equation*}
and $p_j(t)\to0$ otherwise;

\emph{2)} If $u(t)\to \hat{u}$ as $t\to\infty$ then
$$
p_j\to\frac{p_j^0\hat{u}^{d_j}}{\sum\nolimits_i p_i^0\hat{u}^{d_i}}\,,\quad t\to\infty,\quad j=1,\ldots,n;
$$

\emph{3)} If $u(t)\to \infty$ as $t\to\infty$ then
\begin{equation*}
p_j(t)\to\frac{p_j^0}{\sum\nolimits_{i \in J}p_i^0}\quad \mbox{if } j\in J,
\end{equation*}
and $p_j(t)\to0$ otherwise.
\end{proposition}

In Proposition \ref{prop2} we studied the generic case when $E_0[\textbf{b}]\neq 0$. Note that all distributions $\textbf{p}$ such that $E_0[\textbf{b}]=0$ are equilibria of the replicator equation
$$
\frac{d}{dt}p_i(t)=p_i(t)(a_iE_t[\textbf{b}]-E_t[\textbf{a}]E_t[\textbf{b}])\,\quad i=1,\ldots,n.
$$
Therefore, the $(n-2)$-dimensional subset of the simplex $S_n$, $S^0=\{\textbf{p}\colon \textbf{p}\in S_n,\,E_0[\textbf{b}]=0\}$ consists of interior equilibria of the replicator equation. It is worth pointing out that if $u(t)\to\hat{u}\neq 0$ then the limit distribution $\textbf{p}(\infty)$ is an equilibrium belonging to $S^0$. Indeed, in this case $s(t)\to\exp\{-\hat{u}\}<\infty$, while the variable $s(t)$ was defined by the equation $\dot{s}=E_t[\textbf{b}]$, $s(t)=\int_0^t E_{\tau}[\textbf{b}]\,d\tau$. Hence, $s(t)$ is bounded for all $t$ only if $E_t[\textbf{b}]\to 0$ as $t\to\infty$ and $E_0[\textbf{b}]=0$ for the limit distribution~$\textbf{p}$.

Proposition \ref{prop2} not only shows  that the asymptotic states of the replicator equation with the matrix of rank 1 can be only equilibria, it also points out that the case of polymorphic (interior) attracting equilibrium, albeit non-isolated, is not exceptional for the general vectors $\textbf{a}$ and $\textbf{b}$. In particular, if the parameters defined above are such that $\xi_1<0$ and $\xi_2>0$ then the asymptotic state is always polymorphic.

\section{Analysis of a class of replicator equations}
In this section we consider a non-trivial example, where the application of the suggested methods allows us to give a proof for a problem concerning the evolution of sensory systems.
\begin{example} Motivated by a problem in the evolution of sensory systems where gains obtained by improvements in detection are offset by increased costs, \cite{adams2007acc} considered the dynamics of the replicator equations \eqref{S1:Eq:6} with the matrix of the form
\begin{equation}\label{new:1}
a_{ij}=a_ib_j-c_i,\quad i,j=1,\ldots,n,
\end{equation}
where $\textbf{a}=(a_1,\ldots,a_n)^\tau$, $\textbf{b}=(b_1,\ldots,b_n)^\tau$, and $\textbf{c}=(c_1,\ldots,c_n)^\tau$ are given non-negative vectors. They showed, using topological arguments, that in the case of general position (see below) and for dimension $n\leq 5$, the system can have only one global attractor, and this global attractor is an equilibrium having at most two non-zero components. We prove this conjecture, using completely different methods, for an arbitrary $n$, with some additional (mainly technical) conditions on $\textbf{a},\,\textbf{b},\,\textbf{c}$.

Using the notations from Sections \ref{Sec1} and \ref{Sec2}, we obtain the replicator equation in the form
\begin{equation}\label{S3:Eq:5}
\frac{d}{dt}p_i(t)=p_i(t)(a_iE_t[\textbf{b}]-c_i-E_t[{F}]),\quad i=1,\ldots,n,
\end{equation}
and the corresponding selection system is
\begin{equation}\label{S3:Eq:6}
\frac{d}{dt}l_i(t)=l_i(t)(a_iE_t[\textbf{b}]-c_i),\quad i=1,\ldots,n,
\end{equation}
to which the methods from Section \ref{Sec2} can be applied. To clarify the connection with \eqref{S2:Eq:1} we write down explicitly
\begin{equation*}
\begin{split}
\psi_1(\omega)&=\textbf{a}(\omega),\quad v_1(t,H_1)=H_1,\quad H_1=E_t[\textbf{b}],\\
\psi_2(\omega)&=\textbf{c}(\omega),\quad v_2(t,H_2)=-1,
\end{split}
\end{equation*}
where $\omega$ takes the values from the discrete set $\Omega=\{1,\ldots,n\}$. Therefore, the escort system \eqref{S2:Eq:4} has the form
\begin{equation*}
\begin{split}
\frac{d}{dt}s_1(t)&=\frac{E_0[\textbf{b}\exp\{\textbf{a}s_1(t)+\textbf{c}s_2(t)\}]}{E_0[\exp\{\textbf{a}s_1(t)+\textbf{c}s_2(t)\}]}\,,\quad s_1(0)=0,\\
\frac{d}{dt}s_2(t)&=-1,\quad s_2(0)=0,
\end{split}
\end{equation*}
or, integrating the second equation in the last system and dropping index of $s_1(t)$, finally we obtain one differential equation
\begin{equation}\label{S3:Eq:7}
\frac{d}{dt}s(t)=\frac{E_0[\textbf{b}\exp\{\textbf{a}s(t)-\textbf{c}t\}]}{E_0[\exp\{\textbf{a}s(t)-\textbf{c}t\}]}=\frac{\sum\nolimits_{i=1}^nb_ip_i^0\exp\{a_is(t)-c_it\}}{\sum\nolimits_{i=1}^np_i^0\exp\{a_is(t)-c_it\}}\,,\quad s(0)=0,
\end{equation}
where all $p_i^0>0$.

Before analyzing equation \eqref{S3:Eq:7} we note that we consider only the generic case for the game matrix given by \eqref{new:1}. The genericity condition in our case reads as follows: any projections of the three vectors $\textbf{a},\,\textbf{c},$ and $\textbf{1}_n$ to any of the three dimensional subspaces of $\mathbb R^n$ spanned by three standard coordinate vectors are linearly independent (this means that isolated equilibria of the system can have at most two non-zero coordinates, see also \cite{adams2007acc}). Putting in other words, this condition means than for any indexes $i,j,k$ the following holds:
$$
\mbox{det} \left(
  \begin{array}{ccc}
    a_i & a_j & a_k \\
    c_i & c_j & c_k \\
    1 & 1 & 1 \\
  \end{array}
\right)\neq 0,
$$
or
\begin{equation}\label{new:2}
    a_i(c_j-c_k)+a_j(c_k-c_i)+a_k(c_i-c_j)\neq 0.
\end{equation}

We are particularly interested in the limiting behavior of $p_i(t)$ as $t\to \infty$. Here we show, analyzing the escort system of our replicator equation, that $p_i(t)$ tend to an equilibrium of the initial replicator equation as $t\to\infty$ and that this equilibrium is the global attractor of our dynamical system for arbitrary $n$.

First, we make the change of the variables
\begin{equation}\label{new:3}
    \begin{split}
    u&=\exp\{-s\} \Leftrightarrow s=-\ln u,\\
    v&=\exp\{-t\} \Leftrightarrow t=-\ln v.
    \end{split}
\end{equation}
In the new variables equation \eqref{S3:Eq:7} takes the form
\begin{equation}\label{new:4}
    \frac{du}{dv}=\frac{u\sum\nolimits_{i=1}^nb_ip_i^0u^{d_i}v^{c_i}}{v\sum\nolimits_{i=1}^np_i^0u^{d_i}v^{c_i}}\,,
\end{equation}
where $d_i=\max_{i}\{a_i\}-a_i$. Note that at least for one $i$ $d_i=0$. We can also assume, without loss of generality, that $\min_i\{c_i\}=0$ (in general, we scale $\tilde{c}_i=c_i-\min_i\{c_i\}$ and drop the tilde for notational simplicity).

The initial condition for \eqref{new:4} is $u(v=1)=1$.

We rewrite \eqref{new:4} as a dynamical system on the plane
\begin{equation}\label{new:5}
    \begin{split}
        \dot{u}&=u\sum\nolimits_{i=1}^nb_ip_i^0u^{d_i}v^{c_i},\\
        \dot{v} &=v\sum\nolimits_{i=1}^np_i^0u^{d_i}v^{c_i},\\
        u&(0)=v(0)=1,
    \end{split}
\end{equation}
where the derivatives are taken with respect to some dummy ``time'' variable. We remark that system \eqref{new:5} has an isolated equilibrium $O(0,0)$, and the axes $u=0$ and $v=0$ are orbits so that $O(0,0)$ cannot be monodromic (focus or center).  Using the function $f=(u^2+v^2)/2$ we find that
$$
L_tf=\frac{\partial f}{\partial u}\dot u+\frac{\partial f}{\partial v}\dot v=u^2\sum\nolimits_ib_ip_i^0u^{d_i}v^{c_i}+v^2\sum\nolimits_ip_i^0u^{d_i}v^{c_i}>0
$$
for $u>0,\,v>0$. Here $L_t(\cdot)$ is the derivative along the orbits of the dynamical system \eqref{new:5}. The last expression implies that the first quadrant of the phase plane (the one we are actually interested in) of \eqref{new:5} is a repelling parabolic sector (i.e., it is a parabolic sector, see, e.g., \cite{andronov1973qts} for the definitions, for which the origin attracts the orbits when ``time'' tends to $-\infty$).

The qualitative theory of ordinary differential equations on the plane is an extensively researched area (see, e.g., \cite{dumortier2006qtp}), especially in the case when the right hand sides are analytic functions.  In our case this would mean that $d_i$ and $c_i$ are natural numbers, but it is a straightforward procedure for our problem to extend the basic necessary results to the case when $c_i,d_i\in\mathbb Q_+$ (see Appendix), so in the following it is assumed that $c_i,d_i$ are non-negative rational numbers.

If point $O(0,0)$ is not monodromic therefore there are characteristic directions along which the orbits of the dynamical system approach the equilibrium \cite{andronov1973qts}. According to \eqref{new:3}, the behavior of the orbits of \eqref{new:5} when $u,v\to O(0,0)$ determines the asymptotic behavior of $s(t)$ when $t\to\infty$. In the following we shall call the orbits of \eqref{new:3} \textit{$O$-orbits} if $u,v\to O(0,0)$ for positive or negative ``time'' directions. We also recall that
$O$-orbits of system \eqref{new:5} have a power asymptote with a positive exponent $\rho$ and a non-zero coefficient $C$ if
$$
u=Cv^{\rho}(1+o(1)),\quad C\neq 0,\,\rho>0,\,v\to0.
$$

The theory of the power asymptotes of $O$-orbits of an isolated equilibrium in the plane is well-developed (see, e.g., \cite{Berezovskaya2007,berez1975,berezov1976,berez1976,briuno1989lmn}), and we apply it here following mainly \cite{berez1976}.

To summarize the main results we need the notion of the Newton polygon. We assume that all $p_i^0>0$. Introduce rectangular coordinates in the plane and to every $b_i\neq 0$ assign a point $A_i$ with coordinates $(c_i,d_i)$. Consider the convex polygonal line $\mathcal{N}$ passing through points of the set $\{A_i\}$ joining $(0, d_{i_j})$ and $(c_{i_k},0)$ such that each $A_i$ lies above or on $\mathcal{N}$. This line is known as Newton's polygon or Newton's diagram (see Fig. \ref{fig:1}). The Newton polygon consists of a finite number of line segments $\mathcal{N}_j$, whose angles with $x$-axis are between 0 and $\pi/2$, with end vertexes $A_{i_{j}}$ and $A_{i_{j+1}}$, $j=1,\ldots, K-1$, and $K$ is the number of vertexes. To each vertex $A_{i_j}$ of the Newton polygon $\mathcal{N}$ we assign the index of the second type $\beta_j=b_{i_j}$, and to each line segment $\mathcal{N}_j$ we assign the index of the first type $$\alpha_j=\frac{(c_{i_{j+1}}-c_{i_{j}})}{(d_{i_{j}}-d_{i_{j+1}})}=\frac{(c_{i_{j+1}}-c_{i_{j}})}{(a_{i_{j+1}}-a_{i_{j}})}\,.$$
\begin{figure}
\centering
\includegraphics[width=0.45\textwidth]{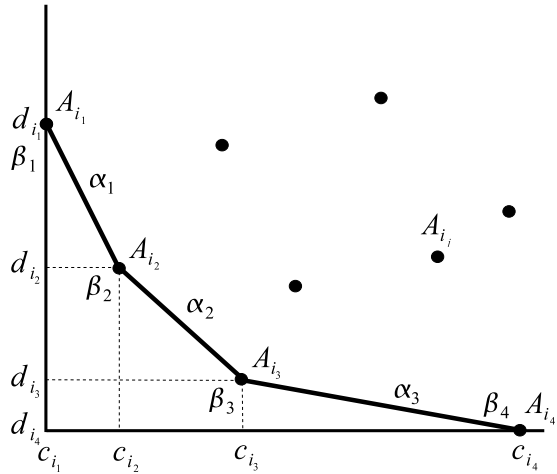}
\caption{The Newton diagram is given by the convex polygonal line passing through the points $A_{i_1},\ldots,A_{i_4}$ (an illustrative example). The diagram consists of 4 vertexes with the four indexes of the second kind $\beta_j$ and of three line segments with the indexes of the first kind $\alpha_j$. See text for details}\label{fig:1}
\end{figure}

Using the results from \cite{berez1976} we have the following theorem:
\begin{theorem}\label{Theor1} Let $d_i,\,c_i\in\mathbb Q_{+}\,,b_j\in \mathbb R_+$, and $d_i\neq d_j,\,c_i\neq c_j$ for any $i\neq j$, $b_i>0$, $p_i^0>0$ for any $i$ in system \eqref{new:5}, and $b_{i_j}\neq b_{i_{j+1}}$ for any two vertexes of line segments of the Newton diagram. Let condition \eqref{new:2} hold, and index of any vertex $A_{i_j}$ be not equal to the indexes of the adjacent edges, i.e., $\beta_j\neq \alpha_j$ and $\beta_j\neq \alpha_{j+1}$. Then all $O$-orbits of system \eqref{new:5} have power asymptotes. The positive exponents $\rho$ of the power asymptotes of $O$-orbits can be

\emph{i)} $\rho=\beta_j=b_{i_j}$, where $\beta_j$ is the index of the second type of the vertexes of the Newton polygon, if for the vertex $A_{i_j}$  $\alpha_j<\beta_j<\alpha_{j+1}$ holds;

\emph{ii)} $\rho=\beta_1$ if $\beta_1<\alpha_1$;

\emph{iii)} $\rho=\beta_K$ if $\beta_K>\alpha_{K-1}$;

\emph{iv)} $\rho=\alpha_j$, where $\alpha_j$ is the index of the first type of the line segment $\mathcal{N}_j$ of the Newton polygon, if the function $\Phi(z)=p_{i_j}^0(b_{i_j}-\alpha_j)z^{d_{i_j}}+p_{i_{j+1}}^0(b_{i_{j+1}}-\alpha_j)z^{d_{i_{j+1}}}$ has non-zero root $z=C$, and this root is the coefficient in the power asymptote.
\end{theorem}

\paragraph{Remark.} The major assumptions in \cite{berez1976} are different from those given in Theorem \ref{Theor1} but follow from them. In particular, for any line segment $\mathcal{N}_i$ it is necessary to consider two functions:
\begin{equation*}
\begin{split}
P_j(u,v)&=p_{i_j}^0u^{d_{i_j}}v^{c_{i_j}+1}+p_{i_{j+1}}^0u^{d_{i_{j+1}}}v^{c_{i_{j+1}}+1},\\
Q_j(u,v)&=b_{i_j}p_{i_j}^0u^{d_{i_j}+1}v^{c_{i_j}}+b_{i_{j+1}}p_{i_{j+1}}^0u^{d_{i_{j+1}}+1}v^{c_{i_{j+1}}},\quad j=1,\ldots,K.\\
\end{split}
\end{equation*}
Then the conditions from \cite{berez1976}, adapted to our problem \eqref{new:5}, can be stated as: the functions $P_j(u,1)$ and $Q_j(u,1)$ cannot have common non-zero real roots, which follows from the fact that $b_{i_{j}}\neq b_{i_{j+1}}$ for any two vertexes of the line segments of the Newton diagram; and the function $\Phi_j(u)=-\alpha_juP_j(u,1)+Q_j(u,1)$ cannot have multiple non-zero real roots, which follows from the fact that $\Phi_j(u)$ is a binomial, and should be identically zero to have multiple roots. The condition for $\Phi_j$, $P_j$, and $Q_j$ to be binomials follows from \eqref{new:2}.

Noting that we are given the initial conditions $u(0)=v(0)=1$ we conclude that there is \textit{only one} orbit passing through the point $(1,1)$. This orbit has a power asymptote $u=Cv^\rho(1+o(1))$ when $v\to0$ (first quadrant is a parabolic sector, where all the orbits have power asymptotes), and the exponent $\rho$ can be either the index of the first type or of the second type of the corresponding Newton polygon $\mathcal{N}$.

Having the power asymptote we can rewrite \eqref{S3:Eq:9} in the variables $u,v$:
\begin{equation}\label{new:6}
    p_i(v)=\frac{p_i^0v^{c_i}u^{d_i}}{\sum\nolimits_{i=1}^np_i^0v^{c_i}u^{d_i}}\,,
\end{equation}
or, using $u=Cv^{\rho}(1+o(1))$,
\begin{equation}\label{new:7}
    p_i(v)=\frac{p_i^0v^{c_i+\rho d_i}C^{d_i}(1+o(1))}{\sum\nolimits_{i=1}^np_i^0v^{c_i+\rho d_i}C^{d_i}(1+o(1))}\,.
\end{equation}

For the following we need (\cite{berez1976})
\begin{lemma}\label{Lemma1}
For any index of the second type $\beta_j$ of the Newton polygon $\mathcal{N}$, which can be an exponent in the power asymptote (see Theorem \ref{Theor1}), we have
\begin{equation}\label{new:8}
    \begin{split}
        d_{i_j}+\beta_jc_{i_j}&=\varepsilon^2_j>0,\\
        d_{k}+\beta_jc_{k}&=\varepsilon^2_j+\tilde{\varepsilon}^2_k,\quad k\neq i_j,
    \end{split}
\end{equation}
where $\tilde{\varepsilon}^2_k>0$.

For any index of the first type $\alpha_j$ of the Newton polygon $\mathcal{N}$ we have
\begin{equation}\label{new:9}
    \begin{split}
        d_{i_j}+\alpha_jc_{i_j}&=\varepsilon^1_j>0,\\
        d_{i_{j+1}}+\alpha_jc_{i_{j+1}}&=\varepsilon^1_j>0,\\
        d_{k}+\alpha_jc_{k}&=\varepsilon^1_j+\tilde{\varepsilon}^1_k,\quad k\neq i_j,\,k\neq i_{j+1},
    \end{split}
\end{equation}
where $\tilde{\varepsilon}^1_k>0$.
\end{lemma}

First suppose, without loss of generality, that $\rho=\beta_j=b_1$. Then, from \eqref{new:7} and Lemma \ref{Lemma1}, we have
\begin{equation}\label{new:10}
    p_1(v)=\frac{p_1^0C^{d_1}(1+o(1))}{p_1^0C^{d_1}(1+o(1))+\sum\nolimits_{i=2}^np_i^0v^{\tilde{\varepsilon}^2_i}C^{d_i}(1+o(1))}\,,
\end{equation}
which yields that $$p_1(v)\to 1, \, p_i(v)\to 0,\,i=2,\ldots,n,\quad \mbox{as } v\to 0.$$

In the case of $\rho=\alpha_j=(c_2-c_1)/(a_2-a_1)$ it follows that
\begin{equation}\label{new:11}
    p_1(v)=\frac{p_1^0C^{d_1}(1+o(1))}{p_1^0C^{d_1}(1+o(1))+p_2^0C^{d_2}(1+o(1))+\sum\nolimits_{i=3}^np_i^0v^{\tilde{\varepsilon}^1_i}C^{d_i}(1+o(1))}\,,
\end{equation}
which gives the limit
\begin{equation}\label{new:12}
    p_1(v)=\frac{p_1^0C^{d_1}}{p_1^0C^{d_1}+p_2^0C^{d_2}}\,,\quad v\to0,
\end{equation}
and a similar expression for $p_2(v)$. Recall that in this case the coefficient $C$ is found as the non-zero solution of
$$
\Phi(z)=p_1^0(b_1-\alpha_j)z^{d_1}+p_2^0(b_2-\alpha_j)z^{d_2}=0,
$$
which finally implies that
$$
p_1(t)=\frac{\alpha_j-b_2}{b_1-b_2},\quad p_2(t)=\frac{\alpha_j-b_1}{b_2-b_1},\quad p_i(t)=0,\quad \mbox{as }t\to\infty,
$$
independently of the initial conditions $p_i^0$.

We have that the $\omega$-limit set of the replicator equation with the matrix \eqref{new:1}, satisfying the genericity condition \eqref{new:2}, consists of the globally attracting equilibrium. This equilibrium is either a vertex of the simplex, and in this case the orbit of \eqref{new:5} passing through $(1,1)$ has a power asymptote with the exponent given by an index of the second type of the corresponding Newton diagram, or this equilibrium is on the $1$-skeleton of the simplex, and in this case the orbit of \eqref{new:5} passing through $(1,1)$ has a power asymptote with the exponent given by an index of the first type of the corresponding Newton diagram:
\begin{theorem}\label{Theor2}
System \eqref{S1:Eq:6} with the matrix \emph{$\textbf{A}$} given by \eqref{new:1}, satisfying genericity condition \eqref{new:2}, and such that the conditions of Theorem \ref{Theor1} hold, always has a global attractor, and this attractor can be only an equilibrium.
\end{theorem}

\paragraph{Remark 1.} According to Theorem \ref{Theor1} we excluded the cases when $\beta_j=\alpha_j$ and $\beta_{j+1}=\alpha_j$. It can be shown (see \cite{berezov1976}) that in this case we have that system \eqref{new:5} can have asymptotes of the form
$$
v=C{v^{\beta_j}}{\ln v}(1+o(1)),
$$
if $\beta_j=\alpha_j$, and
$$
v=\frac{Cv^{\beta_{j+1}}}{\ln v}(1+o(1)),
$$
if $\beta_{j+1}=\alpha_j$. These changes do not influence the limits of \eqref{new:6} and hence the conclusion of Theorem \ref{Theor2} still holds.

\paragraph{Remark 2.} In the original study \cite{adams2007acc} particular ordering of the elements of $\textbf{a}$ and $\textbf{b}$ was used: $0<a_1<a_2<\ldots<a_n$ and $b_1>b_2>\ldots b_n>0$. For the proof given above we do not require any particular ordering of the elements of the vectors $\textbf{a}$ and $\textbf{b}$.
\end{example}

\section{Conclusions}

The replicator equation appears in different problems of evolutionary dynamics in biology and economics; it describes the temporary dynamics of frequencies (or probabilities) in heterogeneous systems under selective force of natural selection, when the fitness itself is frequency-depended.  Typically, only limit sets of the systems are under consideration, in particular the rest points and their characteristics are an usual object to study. The temporary dynamics of frequencies is also of interest and in some applications may be of primary importance. However the problem of studying time-dependent behavior is significantly harder than analysis of the limit sets, especially for systems of high dimension.

In this paper we have presented novel methods to analyze the replicator equation \eqref{S1:Eq:6}. These methods, which potentially can be applied to systems of an arbitrary dimension, are based on the analysis of the corresponding selection system, which should be in particular form  \eqref{S2:Eq:1} (for more details see \cite{karev2009}). In Section \ref{Sec2} we provide an algorithmic approach to find the solution to the selection system assuming that the initial conditions are given. Our approach consists in writing down the corresponding escort system of ordinary differential equations, which in some particular cases can be of significantly smaller dimension then the original one. For instance, in Example  \ref{Examp1} $n$-dimensional system is replaced with one ordinary differential equation.

It is worth pointing out that the suggested approach, in addition to the explicit temporal dynamics, can be used to infer $\omega$-limit set of the original dynamical system; therefore, we only look for attracting asymptotic states and cannot find, e.g., all possible equilibria of the replicator equation. In any respect, $\omega$-limit sets are what usually is of paramount importance in applications because only $\omega$-limit sets are what can be observed from the applied point of view.

For the replicator equation to be suitable for the suggested methods it is usually necessary to apply matrix decompositions briefly described in Section \ref{Sec3}. One of the possible approaches is the singular value decomposition (SVD), which was successfully applied in various static problems to reduce the dimension of the data; here we suggest to use SVD for dynamical problems. Generally, the escort system, whose asymptotic behavior is of particular interest, is $k$-dimensional if the original problem has the matrix of rank $k$. It is therefore tempting to consider an approach when only several singular values are retained, so that we approximate the original matrix of rank $k$ with a matrix of rank $q < k$, for which the escort system if $q$-dimensional.

To illustrate the suggested technique we consider a general replicator equation with the matrix of rank 1 (Section \ref{sec31}). Two major finding are that the $\omega$-limit set is always an equilibrium and that the existence of the polymorphic (non-isolated) equilibria is not an exception for a general matrix $A$ of rank 1.

As an example of the replicator equation with the interaction matrix of rank 2 we consider the problem from \cite{adams2007acc}. In general we show, using the proposed methods and the methods of Newton diagram \cite{berezov1976,berez1976}, that, for arbitrary dimension and under some suitable conditions (see Theorems \ref{Theor1} and \ref{Theor2}), generically one globally stable equilibrium exists on the 1-skeleton of the simplex. We note that our conclusions are based on the studying the limit behavior of the solutions of the replicator equation, which are given in the explicit form, therefore together with the asymptotic behavior the time dependent behavior can be effectively analyzed. It is the next step to apply the presented methods to the replicator equations with the general interaction matrices of rank 2 and to higher dimensional problems.

\appendix

\section{Appendix}

Here we show that, although the main theorems for the asymptotes of the trajectories of the vector field on a plane were proved only for polynomial systems (see \cite{berez1976}), it is straightforward to extend all the results to the system with the right hand sides given by quasipolynomials with rational powers. The result follows from the fact that under some changes of the variables the indexes of the first and second types of the Newton diagram do not change.

Consider the system
\begin{equation}\label{app1}
    \begin{split}
      \dot{u} &= u\sum\nolimits_{i}p_{i}u^{d_i}v^{c_i}, \\
      \dot{v} &= v\sum\nolimits_{i}q_{i}u^{d_i}v^{c_i},
    \end{split}
\end{equation}
where $p_{i},\,q_{i}\in\mathbb R,\, c_i, d_i\in\mathbb Q_+$. We are interested in power asymptotes $u=Cv^{\rho}(1+o(1))$ of the isolated singular point of this system. Let us make the change of the variables:
$$
u=y^m,\quad v=x^r,\quad m,\,r\in\mathbb N.
$$
System \eqref{app1} takes the form
\begin{equation}\label{app2}
    \begin{split}
      \dot{x} &= \frac{1}{r}\sum\nolimits_{i}q_{i}y^{md_i}x^{rc_i},\\
      \dot{y} &= \frac{1}{m}x\sum\nolimits_{i}p_{i}y^{md_i}x^{rc_i}.
    \end{split}
\end{equation}

It is always possible to choose $m$ and $r$ such that the numbers $md_i,\,rc_i$ belong to $\mathbb N$, hence we can apply the technic of Newton's diagram to system \eqref{app2} to find the exponents of the power asymptotes $y=\tilde{C}x^{\tilde{\rho}} (1+o(1))$. The exponents can be only (see the main text and \cite{berez1976}) of the form
$$
\tilde{\alpha}_j=\frac{r}{m}\frac{(c_{i_{j+1}}-c_{i_{j}})}{(d_{i_j}-d_{i_{j+1}})}\,,\quad \tilde{\beta}_j=\frac{r}{m}\frac{p_{i_j}}{q_{i_j}}\,.
$$
Returning to the original variables we obtain that the exponents of the power asymptotes $u=Cv^{\rho}(1+o(1))$ are the indexes of the first or second type
$$
{\alpha}_j=\frac{c_{i_{j+1}}-c_{i_{j}}}{d_{i_j}-d_{i_{j+1}}}\,,\quad {\beta}_j=\frac{p_{i_j}}{q_{i_j}}\,
$$
of the corresponding Newton's diagram built with the rational coordinates $d_i,\,c_i$.

Hence the claim is proved.

Note, that in general we do not need to have identical powers in both equations.

\paragraph{Acknowledgments.} The research of KGP and ASN is supported in part by the Department of Health and Human Services intramural program (NIH, National Library of Medicine).

\bibliography{replicator}

\end{document}